# Variability in electricity consumption by category of consumer: the impact on electricity load profiles

**Philipp Andreas Gunkel[1*], Henrik Klinge Jacobsen[1], Claire-Marie Bergaentzlé[1], Fabian Scheller[1,2]** and **Frits Møller Andersen[1]**

[1] Sektion for Energy Economics and Modelling, DTU Management, Technical University of Denmark, 2800 Kongens Lyngby, Denmark

[2] Faculty of Business and Engineering, University of Applied Sciences Wurzburg-Schweinfurt, Ignaz-Schon-Street 11, 97421 Schweinfurt, Germany

* Correspondence: phgu@dtu.dk; Produktionstorvet 424, 2800 Kongens Lyngby, Denmark

**Abstract.** Residential electrification of transport and heat is changing consumption and its characteristics significantly. Previous studies have demonstrated the impact of socio-techno-economic determinants on residential consumption. However, they fail to capture the distributional characteristics of such consumer groups, which impact network planning and flexibility assessment. Using actual residential electricity consumption profile data for 720,000 households in Denmark, we demonstrate that heat pumps are more likely to influence aggregated peak consumption than electric vehicles. At the same time, other socio-economic factors, such as occupancy, dwelling area and income, show little impact. Comparing the extrapolation of a comprehensive rollout of heat pumps or electric vehicles indicates that the most common consumer category deploying heat pumps has 14% more maximum consumption during peak load hours, 46% more average consumption and twice the higher median compared to households owning an electric vehicle. Electric vehicle show already flexibility with coincidence factors that ranges between 5-15% with a maximum of 17% whereas heat pumps are mostly baseload. The detailed and holistic outcomes of this study support flexibility assessment and grid planning in future studies but also the operation of flexible technologies.

## 1. Introduction

This study provides a comprehensive summary of residential data on electricity consumption at a level of detail that makes it suitable for further studies and applications in research, public services and industry. Electricity consumption is subject to change due to the electrification of heat and transport in the context of the green transition [1]. In response to this development, several key areas in the energy sector, such as the generation of electricity, network planning, grid tariffs and tax design, are being reconsidered [2–4]. Consumer groups with different socio-techno-economic characteristics will therefore face changes in their electricity bills that are more dependent on the timing, location, peak and distribution of their electricity consumption [4]. Distributional cost effects across all socio-techno-economic groups are expected. Consequently, more detailed research on residential electricity consumption, its distributional characteristics and developments is necessary to anticipate future challenges such as network development and policy design.

Residential electricity consumption varies across socio-economic parameters and technical equipment. Previous studies have demonstrated that it is mainly determinants such as the type of dwelling, the heating system used and the charging of electric vehicles that significantly affect consumption levels and daily peaks [5]. Further influences on consumption levels range from the number of occupants [6] to the number of bedrooms [7], the dwelling area [8], the floor area [9], incomes [10] and the household's ownership of physical appliances [11] and electric vehicle [12]. Occupant characteristics and living conditions, which are often correlated with income, also play a significant role [13]. Individual profiles are analyzed by [14] to identify different clusters of consumers based on socio-economic factors determined by survey data. A 3,326 smart meter records dataset is divided into 6 clusters with different peak consumption. Socio-economic factors have large effects on the association with clusters. However, households also showed the characteristic of moving from one cluster to another depending on the season. Therefore, residential patterns regarding similar groups are subject to changing patterns that confirms the need for a detailed and fragmented investigation of residential consumption profiles, both between and within chosen socio-techno-economic groups, since electricity profiles are heterogeneous [15]. Similarly, [16] clusters residential consumption of 5566 households in London to investigate consumption behavior. In the end, the study results in consumer clusters with daily profiles that can help retailers to optimize their market participation based on customer segmentation. [17] improved clustering methods by focusing on behavioral characteristics of consumers changing their consumption pattern over time. [18] further analyzes the change of residential electricity consumption by consumers adopting EV and PV. The authors observed using a difference-in-difference method that demand changes due to new technologies and behavioral changes. Most studies using clustering have a consumption-first focus. However, grid operators, grid and city planners, and policymakers focus on a socio-economic-centric view. Consumer change clusters over time while socio-economic categories are more stable and are relevant to calculate the impact of policy initiatives and infrastructure planning [19,20]. Most research links individual electricity profiles and household characteristics, ranging from analyses of large samples [5] via representative smart-metering surveys [21], longitudinal cross-sectional data [22,23] and specifically collected information on residential groups [6,8–10,13] to analysis of different temporal resolutions [24–26]. The studies cited utilizing common average- and regression-based methods are useful in drawing unified conclusions and average relationships [22,27,28]. However, they deal with various parts of the conditional distribution identically and neglect its heterogeneity [22]. Also, a comparable holistic analysis is missing as studies focus on subsets of categories or specific technologies, which makes comparisons harder due to differences in time, location, temperature or behavior. For specific hours or timeframes, quantile regression is a further option, as applied among others by [15,22,25]. To a certain econometrical degree, those values are useful in understanding consumption. Nevertheless, the unavoidably high level of aggregation makes the data less suitable for estimating the named areas where timing, location, distribution, outliers and noise are expected to be decisive. The investigation of individual profiles is thus necessary due to their heterogeneity [6]. [29] presents a stochastic analysis of plugin and availability pattern of 10 electric vehicle used in the service sector, while [30] analysis the charging behavior of 221 electric vehicle for 78 days in the UK. Plugin pattern of electric vehicle have mostly been analyzed through theoretical and survey approaches such as in [31,32]. Probabilistic heat pump pattern are statistically analyzed by [33] for 19 households for the month of January in Ireland. A main outcome of the study is a coincidence factor for the usage of heat pumps follows a gamma distribution with a strong baseload with long and flat tails relevant for network planning problems. Similarly, [34] shows considerable flexibility potentials provided by energy communities using electric vehicles and heat pumps under uncertainty. However, a clearer view on realistic raw data is required to allow for optimal scheduling of smart home systems that have to deal with uncertainties related customer behavior in relation to charging and heating, but also production from renewables [35–38]. This will help to mitigate and avoid costly distribution grid reinforcement as shown by [39]. Future studies therefore are in need of detailed insights into electricity consumption pattern across and within customer groups including uncertainty to improve prediction and operational models.

To address many of the named shortcomings, this research aims to reveal fundamental differences in residential electricity consumption between socio-techno-economic categories at the individual dwelling level in Denmark. In contrast to previous studies, the analysis does not only focus on average and aggregated consumption analyses. It applies median, variance, and probability approaches and thus the distribution of electricity consumption, which is of great importance to get a sense of the marginal impacts across the individual electricity consumption profiles. While existing articles provide insights about the average relationships of the electricity demand and respective variables, we present distribution shapes of residential electricity consumption profiles for socio-techno-economic consumer groups. This approach offers new insights into significant uncertainties for flexibility measures and grid adaptations at the centers, tails and peaks of the consumption distributions and their temporal appearances. Furthermore, we quantify electricity consumption variations within existing household categories while providing the same general properties such as time, location and climate conditions. In the end, this study provides a holistic and comparable analysis across socio-economic groups and technologies. The analysis is thereby also highlighting the heterogeneity within each group. Consequently, this study provides an analysis of residential electricity consumption that supports grid planning and policy making further and provide a beneficial view for data science and scheduling applications as well as offer direct comparisons between technologies such as heat pumps and electric vehicles that have so far been studied only individually in literature.

Our approach shows that techno-economic determinants like heat pumps influence with higher probability the aggregated peak consumption than electric vehicles while socio-economic determinants such as occupancy, living area, and income show little impact. Although electric vehicles generally contribute with a lower probability to residential consumption than heat pumps, in the event they significantly demonstrate a higher magnitude for individual hours of the day. By demonstrating this, we rely, unlike other studies, not on smart metering surveys and selective field trails from a limited number of households but on a unique and comprehensive dataset from a large number of individual dwellings into account. The dataset covers approximately 720,000 households with an hourly meter of 2017. The Danish Transmission System Operator (TSO) Energinet collects data from all hourly meters and delivers it delivers to Statistics Denmark. The smart-meter data is linked to administrative registers giving reliable information on household categories. Thereby, the applied unique dataset from Statistics Denmark with the large amount of individual household data also decreases the impact of the volunteer bias present in most studies in the literature.

The study is structured as follows. Section 2 introduces the Danish Energy system in general and then presents a closer look on the residential sector as well as the data and material of this study. After that, Section 3 shows the applied methods and measures to compare residential electricity consumption across consumer groups. Section 4 summarizes the main results and compares the influence of different socio-techno characteristics on statistical measures and distributions of residential electricity consumption. The discussion is performed in Section 5. Firstly the results are compared and validated with existing literature and generalized to other geographical locations when possible. In the second part of the discussion the results are put into context of network planning and load forecasting. The last subsection of the discussion outlines research potentials such as grid tariff and tax designs. Section 6 summarizes and concludes the study.

**2. Materials - Residential energy consumption in Denmark**

This section gives at first a general context of the electricity sector in Denmark regarding its generation, consumption and then focuses on the energy demand of Danish households. After that, the case study data is introduced, covering the categorization of household archetypes by socio-techno economic factors. The last subsection summarizes the methodology and presents the calculated indicators.

*2.1. Danish electricity sector and residential consumption*

The Danish electricity sector has undergone a change in production and consumption over the past decades. Since the 1980s Danish energy production shifted towards the expansion of renewable energy resources such as wind power due to the ban of nuclear power originating from strong public resistance [40]. The contribution of wind power to the total electricity production reached already 48% in 2017. Flexible generators, mostly co-generation power plants supplying district heat, covered around 50%, whereas solar PV only played a minor role [41]. Danish politics remain ambitious with their plans to meet the requirements of the Paris agreement and are currently on track [42]. The Danish Energy Agencies foresees an expansion of solar PV by 445% and 643% in 2025 and 2030, respectively, an onshore wind development from 4.4 GW in 2018 to 6.2 GW in 2030, and an offshore wind capacity of 5.6 GW, which represents an increase by 435% [43]. The incoming variable resources cover the rising electricity needs from the electrification of several sectors, particularly the residential, that is also supposed to serve the needed flexibility via demand response.

Unlike several other European countries, Danish electricity consumption is largely affected by residential consumers. Figure 1 summarizes the division of electricity consumption by sector on a typical winter day. The pattern of the industry sector mainly influences the total electricity peak in the morning, including some contributions by the residential sector. Contrarily, the residential sector is clearly responsible for the late afternoon peak.

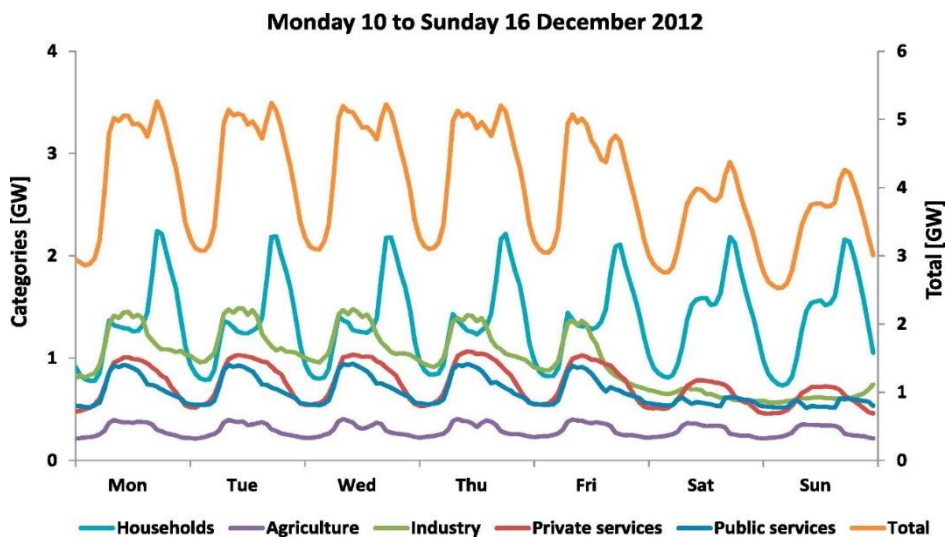
*Figure 1 Danish electricity consumption divided by sectors [26]*

Figure 2 additionally summarizes the Danish electricity consumption and residential energy demand

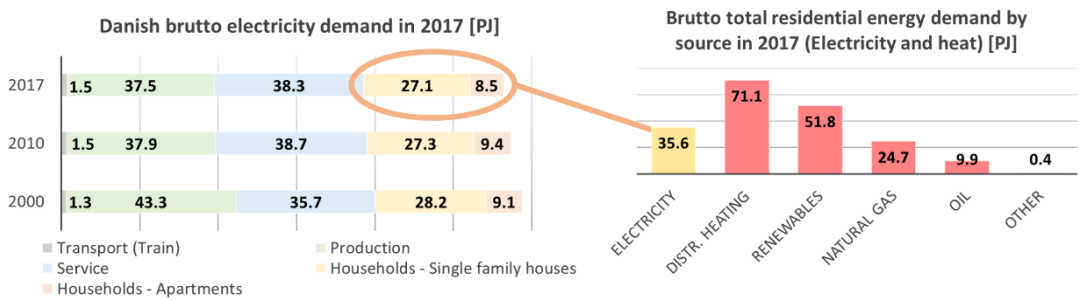
*Figure 2 Summary of Danish electricity consumption in 2017 (left) and residential electricity and heat consumption (right).*

Roughly a third of the national electricity consumption originates from households, which is a higher share compared to other industrialized countries [44]. District heating plays a major role in Danish heating production, covering approximately 68% of the total heat demand, while only around 45% of the total household heat demand comes from centralized sources [41]. Future decarbonization developments impact electricity demand by the decentralized electrification of heat at the expense of emitting natural gas and oil sources. District heating expansion will also take shares in competition, however, only in urbanized areas of the country [43]. Another source of increasing residential electricity demand is coming from the transport sector. Passenger transport in Denmark is responsible for 161.4 PJ, with 66% of it originating from private vehicles. The Danish Energy Agency foresees a significant uptake of private EVs in the 2020s. It is expected that 354,000 out of the 3.35 million vehicles are purely driven by electricity [43]. This 10% of the total vehicle stock will account for 3.6 PJ.

Both heat pumps and electric vehicles will substantially raise domestic electricity demand. While the yearly demand is simple to forecast, the hour-by-hour integration of those technologies is crucial to balance fluctuating electricity generation by wind and solar. A smart integration can lead to substantial benefits in terms of cost and prices, whereas a passive integration imposes threats to a stable electricity system [45]. Thus it is of utmost interest to investigate detailed residential consumption patterns.

*2.2. Danish smart meter electricity data, register data and consumer categorization*

The data were provided by Danmarks Statistik in an anonymized form and were linked to a long list of administrative registers giving detailed information on a number of socio-economic factors, namely citizens, families, households, vehicles, buildings and addresses. The raw data delivered contains around 1.2 terabytes. They are cross-featured, covering approximately four million out of Denmark's 5.6 million citizens. The four million are forming approximately two million households. These households are then cross-featured with the available meter point ID of the smart meters that have already been installed. As the rollout of smart meters in Denmark was still underway in 2017, only around 1.5 million meter points are uniquely connected to households. Lastly, the meter points are connected to the respective electricity consumption profiles for 2017. Due to the ongoing rollout and other disturbances, many profiles do not cover the entire year or have substantial gaps. Therefore, it was decided to sort out meter points whose raw data on electricity consumption have gaps or faulty entries greater than a thousand hours in the corresponding year ending. The data cleaning depends on simplified algorithms. Households that share a meter point with the service sector or agricultural activities, such as farms, are omitted in. Summer houses are also excluded from this study. The accuracy of measurements reaches 0.001 kWh, while the maximum distribution grid connection of households in Denmark is 29 kW most of the time. Consequently, data entries greater than 29 kWh or zero entries including non-numerical values are also flagged as faulty. This study covers in the end approximately 720,000 households and profiles.

Annual total consumption must be corrected for the missing hours in order to harmonize the data and provide comparability. As the holes appear randomly to the greatest extent, a simple approach is also used in this case. The average hourly consumption of the meter point is calculated to round the annual total upwards only: seasonalities are not taken into account. The relevance of this adjustment is limited to Figure B13 and Figure B14. Gaps in the hourly profiles are generally not filled in order to maintain a cleaner and non-altered picture for the distributions of hourly consumption.

For the purposes of this study, the 720,000 households are divided into socio-techno-economic characteristics forming in a total of ninety consumer categories. The characteristics with which to determine consumer categories are selected with reference to three main considerations. The first is to be able to communicate information in as much detail as possible about residential electricity consumption. The second is to provide information with a reasonable level of statistical significance. The third is to be in agreement, at least, with GDPR, but preferably anonymizing the data at an even higher level to cover privacy and ethical concerns. Choosing qualitatively meaningful characteristics underpins two further objectives related to this aim, as well as to future studies. To begin with, the characteristics must be of relevance to residential electricity consumption and variability, which covers both the technical and socio-economic factors. The second objective is related to improving the policy assessment, the re-distribution of cost and the adoption of technologies. Overall, the characteristics and subsequent groups should cover the impact on the variability in and size of household electricity consumption and allow for analyses of policy, financial burdens, and future adoption. We follow the main recommendations suggested by the "Manual for statistics on energy consumption in households"[46] to retain international standards and comparability. The chosen categories also follow the recommendations of [47] and [48] that summarize with a detailed literature review unambiguously connected factors to electricity consumption. Moreover, the categories are also chosen based on an extra literature review to better compare the results to other studies [5,7,8,49].

Table 1 summarizes the chosen socio-economic characteristics and their values. The objective of the chosen characteristics is to show the variabilities and patterns of household electricity consumption.

*Table 1 Chosen socio-economic categories and their respective values.*

| Characteristic name | Characteristics | | | | |
|---|---|---|---|---|---|
| Dwelling type | | AP: Apartment | H: House | | |
| Occupancy | | P1: 1 occupant | P2: 2 occupants | P3: 3-4 occupants | P5+: 5 or more occupants |
| Dwelling area | AP: | A1<$66m^2$ | $66m^2$<A2<$85m^2$ | $85m^2$<A3 | |
| | H: | A1<$110m^2$ | $110m^2$<A2<$146m^2$ | $146m^2$<A3 | |
| Income level | | €1<$240kDKK$ | $240kDKK$<€2<$449kDKK$ | $449kDKK$<€3 | |
| Electric vehicle (EV) | | EV0: No | EV1: Yes | | |
| Heat pump | | HP0: No | HP1: Yes | | |

The consumer categories are divided as follows. Dwelling types are divided into two parameters. The letter *H* indicates that the household resides in a detached or semi-detached house, whereas *AP* stands for apartments. The second characteristic is occupancy, or the number of persons living in the household. P1 and P2 indicate households with respectively one and two residents, P3 contains households with three or four persons living in it, and P5+ represents five or more. The lower third income group represented by *€1* earns up to 240,260 DKK/year, the medium income group *€2* up to 449,097 DKK, and the remaining upper group is included in *€3*. Similarly, households' dwelling areas are divided into three different groups. Since houses and apartments have different size characteristics, the *A1* of houses goes up to 110 sqm, whereas *A1* for apartments covers up to 66 sqm. *A2* for houses goes up to 146 sqm and for apartments up to 85 sqm. The second last characteristic indicates whether electric vehicles are connected to the household (*EV1*) or not (*EV0*). Lastly, the heating source is also determined via a simple binary representation. While *HP0* represents no electrical heating installed, the abbreviation *HP1* indicates that the house or apartment is heated by electricity. Electricity-to-heat installations therefore only cover HPs and exclude electric boilers as the main heating source. Electric boiler are primarily used in summer houses in Denmark. They further do not play a significant role in Danish residential heating and are not preferred as a solution by Danish heat planning. Households connected to district heating are, moreover, only represented in *HP0*, as the heating production of the respective district heat provider was not included in the data. Income and dwelling area levels are divided into two groups using median statistics. In particular, housing type and income are of interest for qualitative assessments, the other characteristics for quantitative calculations.

All possible combinations of characteristics are produced, forming 210 unique groups, ninety of which are ultimately present in this study. The residual 120 groups do not have enough profiles present to fulfill privacy considerations. Additional information on each category is shown in Table A3 by focusing on six chosen consumer categories on which the main focus lies. Households owning both an EV and HPs are excluded from the study due to the limited observations in our dataset, thus limiting the statistical significance and the risk of non-compliance with data privacy.

*3. Methodology*

*3.1 Representing peak hours*

A peak hour is an hour during which the available generation or transmission capacity struggles to meet increasing demand, resulting in additional system costs. There is no fixed definition of when an hour is considered to be the peak, but it can be approximated by referring to the generation mix and the number of hours during the year when peak capacity is called for. On a day-to-day basis, peak periods represented in flexibility programs last from two to twelve hours [50]. The definition of peak consumption varies in the literature. Household consumption largely defines peak consumption in Denmark. Conversely, in other European countries electricity consumption is largely dominated by industry and production. This study uses the top 20% of national gross consumption to represent a large enough spectrum to cover all the characteristics of aggregated peaks. Conventional and controllable central generators mostly use peak capacities within the top 20% of gross national consumption in the period between 2018 and 2021, while it is important to mention that renewables, such as wind dominate electricity production. At the same time, on many occasions, narrower views are applied to cover smaller peak percentages such as the top 5% and top 1% as well. An example of a Danish load duration curve from 2017, highlighting the consumption of a single category, is shown in the appendix Figure B16. The number of peak hours can differ in each month: for example, winter months usually have more peak hours than summer months. Table 2 summarizes the number of hours per month that are defined as peak hours. Accordingly, some months, such as July, do not present any peak hours, whereas January has 321 peak load.

*Table 2 Number of peak hours based on gross consumption in Denmark in 2017.*

| Month | No. Peak hours | Month | No. Peak hours |
|---|---|---|---|
| 1 | 321 | 7 | 0 |
| 2 | 283 | 8 | 7 |
| 3 | 223 | 9 | 62 |
| 4 | 73 | 10 | 209 |
| 5 | 18 | 11 | 294 |
| 6 | 2 | 12 | 255 |

The shown hours in the table will correspond to the index $M$ with the subset $m$ for each month used to calculate means, standard deviations and T- and Welch-Tests presented in the next subsections.

*3.2 Statistical measures and tests*

Additional statistical tests are performed to support the results and qualitative statements of the study. At first, simple measures such as mean and standard deviations are used to provide an overview of the consumption pattern. Both measures are calculated for specific time frames $m$ that correspond to each month shown in Table 2 of the entire set $M$. The mean and standard deviation are also calculated for the entirety of peak hours in the year, neglecting the fragmentation into months. The set $Y$ contains all peak hours in the year, while the view calculations cover the top 20%, 5%, and 1% indicated by $y^{20\%}$, $y^{5\%}$ and $y^{1\%}$ respectively. The mean and standard deviation are calculated according to Equations (1) and (2) for either each month $m$ in the set $M$ or each total yearly peak hours $y$ in $Y$.

$$\overline{q^{m,y}} = \frac{\sum_N q_n^{m,y}}{N} \qquad (1)$$

$$\sigma^{m,y} = \sqrt{\frac{\sum_N (q_n^{m,y} - \overline{q^{m,y}})^2}{N}} \qquad (2)$$

Additionally, statistical tests are used to determine significant differences between the means of two independent socio-economic groups (e.g., comparing the metric scaled consumption of consumers who have a HP or a EV). For comparing the means of two independent groups, the unpaired Student's t-test proves the difference when the data of each socio-economic group are normally distributed and when the variance between the compared socio-economic groups is assumed to be equal. Since the second assumption is not fulfilled in our study, the Welch's test is suggested in the literature, which equips no requirements regarding the variance. Thus we utilized the Welch's t-test as outlined in equation (3) [47] proves the significance of two groups' different means with different variances.

$$t = \frac{\overline{qx_1^y} - \overline{qx_2^y}}{\sqrt{\frac{s^2_{\overline{qx_1^y}} + s^2_{\overline{qx_2^y}}}{}}} \qquad (3)$$

$\overline{qx_1^y}$ and $\overline{qx_2^y}$ represent the average of the random picks of both groups and their respective standard errors $s^2_{\overline{qx_1^y}}$ and $s^2_{\overline{qx_2^y}}$. Random picks from the distribution of both groups are taken according to the amount of household count in the original data (see Table A3 in the appendix). The test has been repeated 50 times for the same hypothesis and tested against two groups. The tests are performed for the three sets of $Y$ to prove that the populations are different and look into more narrow definitions of peak hours. For simplicity reasons, only the average of the 50 means of the random picks and the percentage of acceptance of the Null-hypothesis of the 50 tests are communicated in this study. The main objective is to prove that the mean of households with and without HP and EV are significantly different. The Welch tests are performed to support statements and the general exploration of the data. We test the hypothesis that two consumption profiles, without HP and EV compared with HP or EV, have equal means. Moreover, we further test if the means of consumers with HP are equal to those of consumers with EV. A rejection of the hypothesis proves that they are significantly different.

*3.3 Probability distribution of individual consumption and charging*

This study further dives into detailed views of probability distributions within specific socio-techno-economic categories. The figures in the result section include probability distributions, mean, median, and standard deviations. Thus the calculation of those values contains consumption inputs of the specific hour independent of peak and off-peak hours. Looking into individual hours and patterns is supposed to give a better context for the aggregated main outcomes.

Additionally, the coincidence factor of EV charging is further approximated. The coincidence factor is defined as the probability of simultaneous charging [51]. Smart meters in Denmark do not meter EV charging or heat pump consumption separately. While the usage of heat pumps is more difficult to distinguish from traditional consumption, EV charging shows a clearer pattern due to the installed charger capacities being able to consume a multitude of traditional residential consumption. Thus the event of charging can be approximated assuming that surpassing a specific total consumption level is connected to a likelihood when comparing identical socio-techno-economic categories with the only difference of owning EV or not.

**4. Results**

Section 4.1 shows the load distributions enabling the differences in the median and variability of the consumption to be visualized. In section 4.2 the profiles are generalized for the top 20% of yearly Danish peak hours in 2017 to quantify the impact of dwelling area, EVs and HPs on residential demand in Denmark. Section 4.3 extrapolates the impact on the load duration curve to anticipate future developments and load effects. While sections 4.1 and 4.2 focus on peaks in consumer categories, section 4.3 uses aggregated consumption and peaks.

*4.1. The impact of consumer categories on peak consumption*

Monthly averages and standard deviations of peak consumption are shown in Figure 3. Peak hours are defined as the top 20% of hours of gross consumption. The color of the tiles indicates the monthly average consumption of each consumer category during peak hours. The degree of variability, which is also indicative of the distribution, is shown as the standard deviation of peak-hour consumption in each tile for each month.

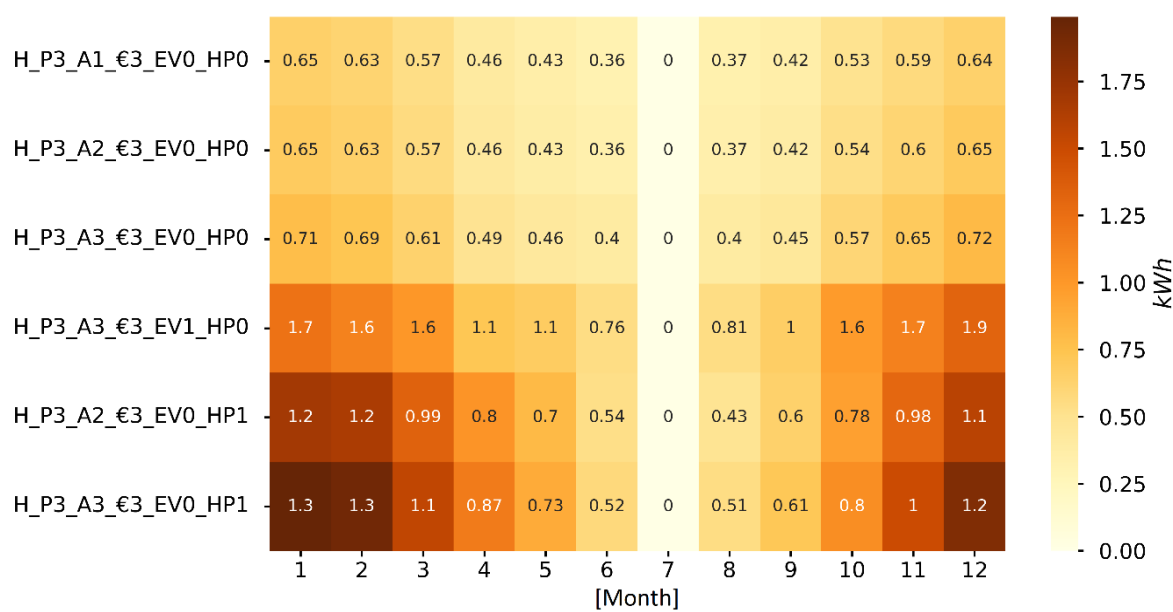

*Figure 3 Monthly average and standard deviation of residential electricity consumption during the top 20% peak hours of the year. The standard deviation of the consumption is represented by the value within each tile in kWh. The categories*

*share three to four person occupancy (P3), house (H) as housing type, and belong to the high-income group (€3). The first three categories starting from the top increase dwelling area (A1, A2, and A3), while the household with a large dwelling area has an EV (EV1). The two last categories are houses with heat pumps (H1) with medium and high dwelling areas (A2 and A3) and no EV (EV0).*

The first three rows in Figure 3 contain the same socio-economic categories with rising dwelling areas from A1 to A3. The winter months show a naturally higher average consumption during peaks than the summer months. The effect of income is comparably small, as shown in Figure B15 in Appendix B, presenting the complete view of all categories. The occupancy figure generally increases average consumption during peak hours, as well as across the entire year (see Table B4 in Appendix B).

Adding an EV to a household (EV1) with a large area imposes a considerable jump in average consumption during peak hours. EVs increase the average peak consumption by approximately 65% in some of the most critical hours and months. At the same time, what is most striking is the degree of variability. The maximum standard deviation without an EV is around 0.72 $kWh$, whereas vehicle charging pushes the standard deviation up to 1.9 $kWh$. Only a certain probability is attributed to home charging events. Not all vehicles have to charge every day, and consequently a considerable share of EV owners follow the pattern of the original socio-economic group. Households that charge their vehicles have particularly high consumption, thus driving up the standard deviation and skewing the distribution.

The last two consumer categories in Figure 3 add HPs (HP1) to medium and high dwelling areas households. Average consumption during peak hours is by far the highest in this comparison. 1.69 $kWh$ and 1.99 kWh of average consumption in January resulted in the largest pressure on the energy system. In contrast, the highest detected standard deviation is comparatively low at 1.2 $kWh$ and 1.3 $kWh$ for medium and large dwelling areas respectively.

Both measures confirm the tendencies that HPs are more likely to contribute to the aggregated peak. The comparison of first the households without EV and HP with the two respective groups with both technologies resulted intuitively that adding one of both technologies adds significantly to the mean (see Table B5 and Table B6 in Appendix B). Furthermore, Welch's t-test also provided evidence that the mean of households with HP are significantly higher and different to the households owning only EV.

*4.2. Anticipating future challenges on load duration curves*

EVs and HPs connected to households have considerably different consumption characteristics during peak hours. In particular, EVs are at the center of attention in the local energy system and grid planning [52].

Consequently, the next figures extrapolate an entire consumer category purchasing an EV or HP. We use the entire load profile of households owning an EV or a HP, and scale them to all 54,445 households adopting EVs and HPs, respectively, by exchanging and scaling the load profiles. This results in different load duration curves and timings of peak consumption. Thus, Figure 4 and Figure 5 quantify the impact of an entire consumer category purchasing an EV or HP on national loads with a special focus on the top 20% with a reordered curve. The household type used as a basis remains unchanged. The load duration curve for 2017, including the share of the consumer category, is shown in the appendix, Figure B16.

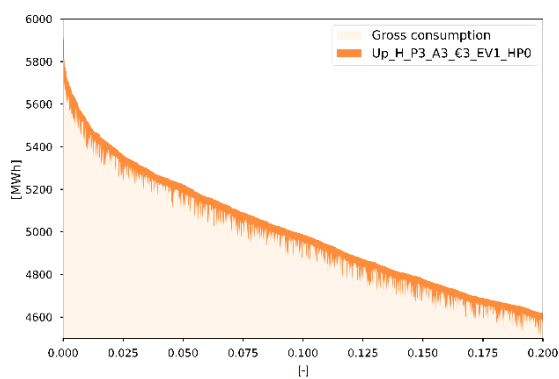

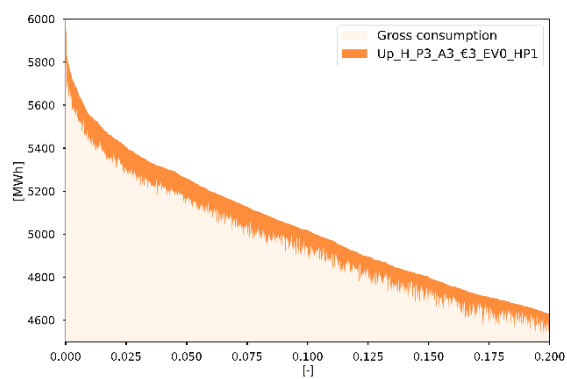

*Figure 4 Top 20% of the Danish load duration curve of 2017 when the entire consumer category of a three to four person household living (P3) in a house without a heat pump (H0) located in the high-income (€3) group and large dwelling area (A3) with 54,445 single households buys an EV. The extrapolated (Up) curve converts to "Up_H_P3_A3_€3_EV1_HP0".*

*Figure 5 Top 20% of the Danish load duration curve of 2017 when the entire consumer category of a three to four person household living (P3) in a house without a heat pump (H0) located in the high-income (€3) group and large dwelling area (A3) with 54,445 single households buys an EV. The extrapolated (Up) curve converts to "Up_H_P3_A3_€3_EV0_HP1".*

The addition of an EV does not contribute as much to peak hours as the addition of an HP in this consumer category. The degree of variability in consumption with EVs is a further evidence that hours may have both higher and lower charging requirements. However, HPs consistently contribute to the peak through their consumption. In particular, the last 10% of the load duration curve shows consistently high consumption. The maximum consumption of the EV group during peak hours yields 140 $MWh$, whereas HPs contribute to the peak with up to 160 $MWh$. The average consumption follows the same intuition at 61 $MWh$ and 89 $MWh$ respectively, whereas the median is 44 $MWh$ and 89 $MWh$. This means that the aggregated residential HPs cause 46% higher average consumption during peak hours and a 14% higher maximum aggregated peak. The means are significantly different, as proven by Welch's t-test in Tables B4 and B5 in Appendix B for different peak level definitions of the top 20%, 5%, and top 1%.

The consumption profile of HPs also raises concerns when the entire load duration curve in the appendix, Figure B18, is compared to the EVs in Figure B17. The highest consumption by HPs comes during the peak period. Conversely, EV consumption is not significantly greater during peak hours compared to non-peak hours (see also Figure B16). Specific consideration of the flexibility potential is therefore needed. Peak hours are usually located in winter periods, and HPs have to supply heat close to their capacity limits. The ability to shift energy is also limited without the addition of thermal heat storage, while most HPs are not likely to have been used flexibly in Denmark in 2017.

The case for EVs might not be as simple. Afternoon peaks through charging are still dominant, as the profile in Figure 8 will show in the following subsection, but increased charging in the later hours or early morning hours also occurs. The median of the highest recorded consumption of the EV category is around 11.65 $kWh$. Around 1% of household profiles had a maximum consumption of 25 kWh, while 98% of households owning an EV stay below 16 $kWh$ over the entire year. Overall two main clusters within the consumer category are visible, one with a maximum consumption of 5-7.5 $kWh$, the other of 11-15 $kWh$.

EV and HP will have an impact on the Danish aggregated load in the future. However, individual consumption and the contribution of each household are also important for several stakeholders, which is further reviewed in the next subsection.

*4.2.1 Contribution of individual consumption pattern to the aggregated load curves*

The load distribution of four consumer categories is presented for 5[th] January 2017, which is a peak load day in Denmark. As a reminder, the distribution aggregates the consumption of a multitude of individual households that day. Individual households can therefore have a significantly different pattern while still being present in the distribution. Color shades represent a 5% probability band. In contrast, the solid line is the median consumption, the dashed line is the average consumption, and the diamond is the standard consumption deviation during the hour.

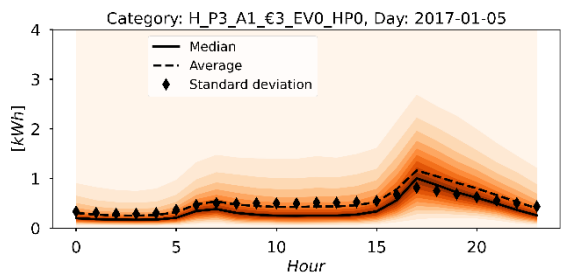

*Figure 6 Load profile distribution of a three to four person household (P3) living in a house without a heat pump (H0), nor an EV (EV0), located in the high-income group (€3) and a small dwelling area (A1).*

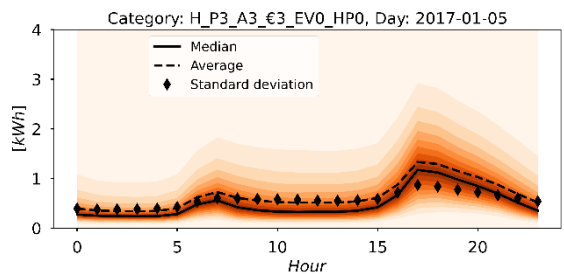

*Figure 7 Load profile distribution of a three to four person household living (P3) in a house without a heat pump (H0), nor an EV (EV0), located in the high-income (€3) group and large dwelling area (A3).*

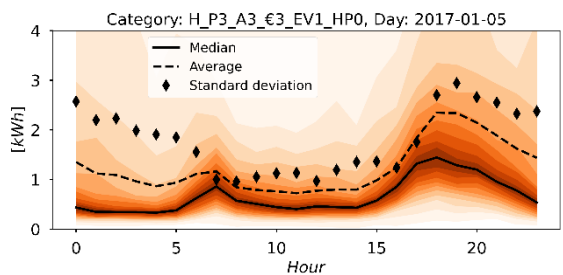

*Figure 8 Load profile distribution of a three to four person household living (P3) in a house within the high-income (€3) and area group. The household owns an EV (EV1) but no HP (HP0).*

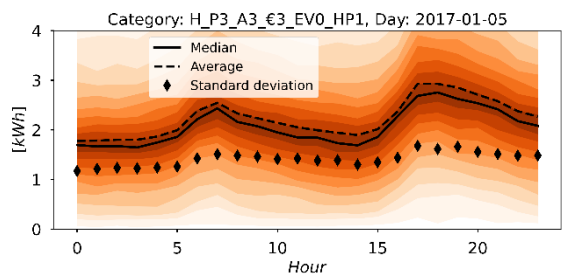

*Figure 9 Load profile distribution of a three to four person household living (P3) in a house within the high-income (€3) and area group. The household owns a heat pump (HP1) but no EV (EV0).*

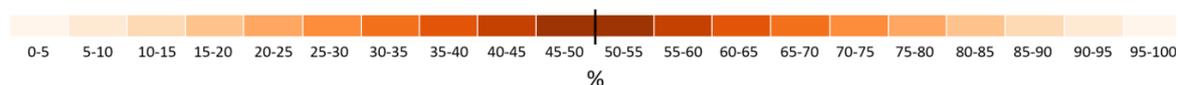

Figure 6 and Figure 7 show load profile distributions for the household type living in small (upper left) and large (upper right) dwelling areas, without HPs or EVs.

Both profiles show known characteristics of residential electricity profiles, including a small peak in the morning and a large one from 5 to 6 p.m. While the house with the small dwelling area has a median consumption of around 0.998 $kWh$, the large area house has a 14% higher median. 95% of the residential consumption in large houses stays below 2.94 $kWh$ from 5 to 6 p.m., which is 10% higher than the small area house with 2.67$kWh$. A large dwelling area and home occupancy are significant drivers of consumption across almost all combinations of categories. Overnight, demand always falls to similar levels, regardless of the occupation rate or size. The category 'income' does not show any evidence of specific and recurrent impacts across all categories in Denmark (compare, e.g., Table B4 or Figure B13 with Figure B14 in Appendix B).

Figure 8 presents the distribution of the electricity demand when adding an EV to the same socio-economic group as in Figure 7, while Figure 9 adds an HP to the group.

Adding an EV to the household (lower left figure) introduces a strong and more chaotic term. This is partially due to the lower number of households within the category, but more importantly due to the addition of a significant source of demand. The largest consumption detected on this specific day occurred from 1 a.m. to 2 a.m. at around 15 $kWh$ (see Figure B11 in Appendix B). The variability increases significantly and spikes in the late afternoon hours compared to Figure 7, then consumption falls overnight and has a second rise shortly before the morning hours. Around 95% of households with an EV stay below 5.1 kWh that day, compared to 2.94 kWh in households without an EV at 5 to 6 p.m. The largest standard deviation in the category with an EV occurs at 7 to 8 p.m. at around 2.95 $kWh$, while the median lies at around 1.27 $kWh$, and the average at 2.3 $kWh$. This means that most households are likely not charging their EVs that day. At the same time, significantly higher consumption is detected in the event of charging that has a likelihood of 14-18% in the afternoon hours. Consequently, the consumer category that owns an EV tends to have a skewed distribution characterized by a high standard deviation and a noticeably lower median than average consumption. This distributional pattern can also be seen in the load duration curve in Figure 4. A smaller base consumption is mixed with peaky spikes following the statistical behavior of a positively skewed distribution.

The same household with an HP instead of an EV shows a different distribution (see Figure 8). The overarching pattern is an increase in consumption compared to Figure 7 and a symmetrical cloud around the median. At the same time, the cloud does not form diffuse spikes as for households with an EV. The highest median and average consumption are present at 6 to 7 p.m. at 2.93 $kWh$ and 2.75 $kWh$ respectively. The degree of variability is relatively low compared to the pattern for EVs, which has a standard deviation of only 1.62 $kWh$. HPs increase consumption, and the profile is more consistent with a well-defined distribution. Extreme spiky events occur less often, but the overall level of consumption is considerably larger with a high probability which is further presented in then next subsection. Flexible consumption and smart operation are not clearly visible. With a lower standard deviation than the EV option and a median that is comparable with the average, the distribution can be characterized as tending to be symmetrical. This can also be acknowledged by reviewing the load duration curve in Figure 5. In contrast to the EV, the inclusion of HP adds a higher mean consumption and only moderate spikes following the characteristics of a symmetrical distribution.

These examples of day and consumer categories highlight the influence of the characteristics on the median, average and standard deviations of the consumption rate, as do the fundamentally different profile shapes and probabilities of consumption when HPs or EVs are part of a household. Because of the characteristic differences between a household with and without an EV, the charging habits can be further viewed in detail in the following subsection.

*4.2.2 Residential electric vehicle charging – the coincidence factor*

In connection with the load duration curve, the coincidence factor is becoming more relevant for grid and energy companies. The coincidence factor is the probability of vehicles charging at the same time. An approximation is visualized in Figure 10. EV charging can not be directly withdrawn from the data set as the household and charging consumption are metered together. At the same time, comparing houses with an occupancy of 3-4, high dwelling area and income with and without EV, the consumption profiles give strong indications. The average probability of the given household without an EV consuming more than 3 *kWh* is 0.5%, whereas 4 *kWh* or more reduces the probability to 0.1% for the entire year. Therefore, we assume that the likelihood of charging is significant enough when the households with EV are consuming more than 3 *kWh* or 4 *kWh* per hour. The smallest charger capacities that are sold are usually around 3.7 kW. The light orange line in Figure 10 shows the probability of a household to consumer more than 3 *kWh* per hour and the dark orange line with more than 4 *kWh* per hour.

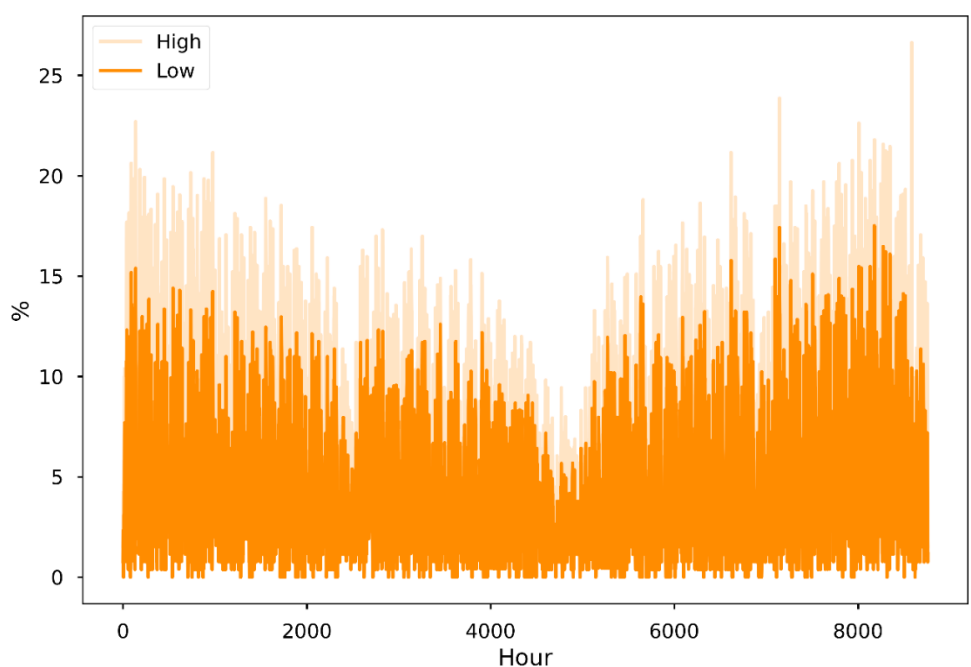

*Figure 10 Bandwidth that shows the probability span of home charging. The light orange "High" line shows the probability of consumption above 3 kWh or more, and the dark orange "Low" line shows the probability of 4 kWh or more of consumption by the category "H_P3_A3_€3_EV1_HP0".*

It is clearly visible that EVs are usually not charging at the same time. The probability exceeds only once a quarter of the entire fleet. This outcome aligns very well with similar studies that are investigating theoretical approaches based on travel surveys on coincidence factors in the Danish case[51]. Looking at 4 *kWh* consumption per hour reduces the coincidence factor even to a maximum 17%. The probability of charging is further underlays seasonal differences, likely due to increased consumption from modal choices and generally higher consumption during colder weather. Summer holidays are also visible, showing lower coincidence factors. At the time of the study, neither specific conclusions on charging stations along highways nor higher consumption in summer houses can be given. Generally between 5% to 15% of the Danish EV fleet is charging at the same time in the afternoon hours.

All in all, the effect of residential HPs on aggregated consumption is larger, while EVs have a greater impact on the peak consumption of individual households. However, EV have a more limited impact on the aggregated profiles with probabilities of a maximum around 25% to charge at the same time.

**5. Discussion**

The discussion section is divided into three subsections. At first, the presented results are compared with existing literature as well as countries and thereafter generalized. The following subsection highlights the impact of our results on improving grid planning and quantifying flexibility. The last subsection shines a light on the potential design of new grid tariffs and their redistributive effects on individual households.

*5.1 Comparison of outcomes to literature and genererlization*

Our methodological choice to focus on distribution characteristics of residential electricity consumption curves as well as on median, average and standard deviations shows that including these measures to anticipate the challenges caused by the integration of EVs and HPs in households is essential. Our results support consumption forecasting, as well as grid and generation planning because we provide a holistic view on profiles that allow for direct comparisons diminishing volunteer bias effects, different times, locations and climate zones when comparing several studies. We demonstrate that HPs are three times more likely to consume during peak periods than EVs for representative consumer categories.

The chosen methodology and presentation of the electricity data in particular brings value to understanding pattern between categories and technologies and reveal further characteristics. However, the outcomes still allow for the comparison and validation with former studies. Residential consumption is generally characterized by two consumption peaks in the morning and evening hours. Our findings align with comparable studies showing similar peak effects in Denmark [5] and other countries like Ireland [14] and Norway [53]. In particular, in households with an occupancy of 3-4 in Figure 6 patterns follow the aggregated findings of [5]. However, it has to be acknowledged that the shown distributions are aggregated by category in the compared study, simultaneously disregarding that some households contribute differently to the probability distribution. A change in behavior of households belonging to the same socio-economic category has been demonstrated by [14] and is likely also present in our data. Households moved with their consumption pattern across different clusters in a year's timespan. Comparing individual peaks of Danish and Irish households further shows a different afternoon/evening peak timing, suggesting a 1-2 hour later incident for Irish households [14]. Similar results are achieved by [54] also using clustering approaches to unravel socio-economic effects on residential consumption in the case of Pecan Street, TX, USA.

Residential consumption in China is much different to Danish in terms of total daily consumption ranging between 6.98 and 53 kWh and variability with an average daily standard deviation of up to 6.11 kWh [55]. At the same time, the socio-techno-economic characteristics of the high values are unknown, which diminishes the comparability of both studies. Large detached houses' yearly consumption with heat pumps comes closer to a study conducted in Norway [53]. A high share of electric boiler, heat pumps and electric radiator characterizes residential heating in Norway. [53] concludes a yearly consumption that is considerably higher than in Denmark. A possible explanation are differences in climate and the fact that the present study only considers heat pumps whereas the studied case in [53] also considers less efficient electric boiler. Similar studies conducted in Denmark show further similar flat hour-by-hour profiles due to electric heating [5]. The symmetric distribution of HP usage is also seen by [33] that has fitted the coincidence factor for HP into a gamma distribution in an Irish test case. The gamma distribution suggest the characteristic of a baseload pattern also seen in Figure 9 followed by a long tail that indicates high peaks during extreme cold days that could also be seen in the top 1% of consumption showing the highest mean. This shows, that climate conditions influence the relative peak importance of HP demand categories. Regions with similar or colder climate than Denmark will have a higher peak influence from HP, but milder climate would have to consider also the high variation in load distribution characterizing the EV demand.

The probabilities for EV charging identified in our study are in line with the lower probability ranges identified in theoretical calculations for Denmark. [51] calculated the coincidence factor dependent on the travel surveys and the maximum charger capacity. The probability of charging reduces with the number of vehicles considered to calculate and the charger's size resulting in a shorter charging time. Calculating the coincidence factor for 100 EV results in a factor of 38% with a 3.7 kW charger, 20% for an 11 kW charger, approximately 14% for a 22 kW charger, and total of 24kWh in charging demand. Our study looks into the raw data of 265 EV (see Table A3 in the appendix) and results in a likely maximum coincidence factor of around 17%. These outcomes show that the given theoretical range by [51] covers the empirically seen coincidence factor. It however has to highlighted that our empirical data even suggests a much lower range coincidence factor between only 5%-15% (see Figure 10). The total energy demand by the EV remains unknown in this study which also have an influence on the outcome and suggest that the average demand seems lower than in the theoretical calculations. The total energy demand is further analyzed in a similar empirical study on the coincidence factor in the UK by [30]. The charging behavior of smaller EV was analyzed and resulted in higher coincidence factors of around 36% for 100 EV. As the battery sizes are given with 24 kWh and the charger capacity lays mostly at 3.6kW, the higher coincidence factor is following theory and allows for the assumption that Danish EV owner have larger battery and charger capacity reducing the coincidence factor to the UK study. The seasonal variation of EV charging is also analyzed in more detail in [12], concluding a similar seasonal pattern with lower demand in the summer and higher in the winter. At the same time, the charging demand is aggregated to averages, thereby flattening the residential consumption curves. Moreover, all hours have been merged, whereas our study, in particular, focuses on peaks. Thus [12] illustrates that February and March can have higher average residential consumption than December in the late afternoon and evening hours. However, the higher consumption happens less during peak hours (compare with Figure 3). Even though both studies consider similar data, the presentation and analysis of both lead to slightly different conclusions that can be attributed to the used methodologies. Further, the lower coincidence factor for EV in Denmark can have a strong influence on grid planning and policy decisions.

Our findings on the consumption profile of specific groups of households can be generalized to countries with comparable income levels, forms of dwelling or climate, such as certain north and west European countries. EV consumption depends on factors like driving patterns, and climate [32,56,57]. Nevertheless, due to similar conditions, it can be expected that the fundamental distribution characteristics are similar in countries neighboring Denmark. Similarly, the outcomes of this study regarding HPs can be generalized to a certain degree because they are dependent on the presence of heat storage or insulation standards. HP consumption pattern are generally technology-specific and likely to be transferable to other countries while taking into account climate and behavioral differences [58]. The low coincidence factor for EV should encourage consideration of new types of incentives driven by policies. This argument is all the more important, as HPs are a key component in efforts to decarbonize the heating sector in many countries, particularly in those relying predominantly on fossil fuel-based domestic heating, such as Germany, the United Kingdom, the Netherlands, Poland [59] or the United States [60].

Differences in incomes have their greatest impact on the probability that new technologies such as EVs and HPs will be adopted. Subsequently, the effect of income shows itself when different categories are created for example in choice of dwelling size. Similarly, the effect of dwelling areas on electricity consumption becomes more significant when HPs are deployed (see Figure 3). In our study, income does not show a recurrent effect on electricity consumption comparing dwelling types and areas. The average, median, and standard deviation differ across groups but without a visible pattern that allows drawing verifiable conclusions. This is in contrast to other regions in the world with different income distributions [61], [62]. Subsection C.1. in the appendix compares the outcomes for Denmark to other countries and studies but leaves the possibility open for further research.

*5.2 The impact of the results on grid planning and the identification of flexibility options*

Our method and results improve knowledge about demand level and profile variation for consumer categories, which enable better demand projections at the local and national levels. Grid planning and identification of flexible demand components can benefit from this, supporting cost reductions and electrification of household demand at the least cost. Utilizing the flexibility of HPs in local grids through local incentives and control equipment requires that largely expected potentials can be identified well in advance, allowing investment in control and the establishment of incentive schemes. Our observation of heterogeneous consumption within each group reveals uncertainty margins and variabilities, which gives distribution system operators in-depth knowledge of the risk of reaching capacity limits in the grid as residential demand increases due to HP installation and EV purchase. Our empirical coincidence factor for EV charging could be used such as in [51] for grid planning and the effect compared to the ones found for UK in [30]. For optimal scheduling analysis of residential electricity demand systems like in [35] we provide details of demand profiles and variation in the composition of household categories with effect on loads.

With the expected changes in the composition of household demand, the variation in demand level and load profiles may increase in the future as EV's and HP's will influence the aggregate load profile increasingly, as studied in [60]. Moreover, the consumption categories are likely more related to income concerning the level of demand due to different technology adoption rates. Concerning flexibility potentials, a relevant study can address whether the EV's and HP's effect on load profiles for households will result in more flexibility in high-income households than low-income households. If this is the case, low-income households may be more exposed to higher variation in electricity prices and, in particular, a change of grid tariffs to more dynamic tariffs.

*5.3 Potentials in innovating grid tariff and tax designs and simulating distributional impacts using individual demand data*

Design of network tariffs reflecting the grid cost impact of demand components is in focus in many countries, but the fairness of tariffs is also of serious concern. Our methodology and data make it possible to examine tariff and tax changes, including the distributional impact on various consumer categories, in even more detail than [4]. Especially, a grid tariff design that incentivizes grid-friendly behavior while not harming poor households can be considered further. The probabilistic nature of consumption, etc., here plays a vital role, as shown by [63]. Our use of microdata provides additional knowledge into how the presence of EVs and HPs in households affects tariff payments [63].

This can allow for investigations on improved economic signals for households to change consumption patterns and install flexibility controls following network constraints while preserving the aggregate tariff revenue for local grid operators. Specifically, new tariff schemes such as Time-Of-Use (TOU), Critical Peak Pricing (CPP), and other dynamic tariffs are currently a significant subject in research and vital for grid operators to limit the occurrence and impact of congestions. According to our results, peak-based grid tariff designs will likely burden heat pump owners more than electric vehicles'. A common assumption has been that EVs threaten peak capacity limits and the subsequent trigger for grid expansion needs. Our data suggest that heat pumps contribute more to the top 20% and also the top 10% of peak consumption in Denmark (see Figure 4 and Figure 5).

For the combined analyses of taxes and network tariffs, our data allows analyzing how the energy-poor can be protected from paying disproportional higher shares of taxes and network costs. Our methodology includes variation among and within socio-economic categories also depending on technology use, particularly if dynamic taxes/tariffs or taxes specific to heating or charging consumption are considered.

**6. Conclusion**

This study gives a detailed view of residential electricity consumption disaggregated by several consumer categories. It aims to provide a holistic as well as a comparative investigation based on socio-economic and technological characteristics. By highlighting further the heterogeneity within consumer categories, we provide grid operators, network and city planners, and policymakers insights that can be used to improve infrastructure planning and policy initiatives by contributing easily usable categorizations. Electricity consumption of an extensive dataset covering 720.000 Danish households of 2017 is thus analysed. Overall, heat pumps and electric vehicles increase consumption significantly, whereas other characteristics such as occupancy and dwelling size are less impactful. Cross-correlations exist between the level consumption, dwelling size and the existence of heat pumps and further between income and the adoption level of heat and transport technologies.

While individual household profiles with electric heat pumps are characterized by a nearly symmetric distribution across all hours of the entire day with a high average and median consumption with low standard deviation, household profiles with electric vehicles connected to home chargers show a positively skewed distribution in most of the hours with lower average and median consumption than heat pump profiles but with a significantly higher standard deviation (median closer to the lower or bottom quartile).

The analysis of the top 20% of Danish gross consumption reveals that heat pumps influence with higher certainty the peak consumption than electric vehicles, while other socio-economic factors such as occupancy, living area and income show lower impact. The standard deviation of electricity consumption by households with electric vehicles across the entire year is two to three times larger than with heat pumps, but at the same time, the average residential consumption is by 5-70% lower depending on the season. Even in the summer months heat pumps show a higher average consumption.

At the same time, other socio-economic factors, such as occupancy, dwelling area and income, show little impact. Comparing the extrapolation of a comprehensive rollout of heat pumps or electric vehicles indicates that the most common consumer category deploying heat pumps has 14% more maximum consumption during peak load hours on the load duration curve. Heat pumps also impact with 46% more average consumption and a twice the higher median compared to households owning an electric vehicle. The probability of charging of electric vehicles indicated by the coincidence factor shows maximum levels of 25% in Denmark in 2017. Seasonality effects are detected with higher factors in the winter times and the lowest during summer holidays.

This study provides several outcomes that can be generalized to neighboring countries or industrialized countries in similar climate zones. In particular, the distribution characteristics of heat pumps and electric vehicles in households are helpful to anticipate challenges in terms of grid and generation planning as well as policy design.

**Author contributions.**

Philipp Andreas Gunkel: Conceptualization, methodology, software, analysis, writing

Henrik Klinge Jacobsen: Methodology, writing, supervision

Claire-Marie Bergaentzlé: Conceptualization, writing, supervision

Fabian Scheller: Conceptualization, methodology, writing, supervision

Frits Møller Andersen: Methodology, writing, supervision

All authors have read and agreed to the published version of the manuscript.

**Funding.** This article was partly funded by the FlexSUS project (nbr. 91352) that received funding in the framework of the joint programming initiative ERA-Net RegSus, with support from the European Union's Horizon 2020 research and innovation programme under grant agreement No 775970. Fabian Scheller kindly acknowledges the financial support of the European Union's Horizon 2020 research and innovation programme under the Marie Sklodowska-Curie grant agreement no. 713683 (COFUNDfellowsDTU)

**Conflicts of interest.** The authors declare no conflict of interest. The funders had no role in the design of the study, in the collection, analyses, or interpretation of data, in the writing of the manuscript, or in the decision to publish the results.

**Appendix A**

*A.1. Additional characteristics of a defined subset of consumer categories*

*Table A3 Additional socio-economic characteristics of each category in the available dataset. P3! Indicates the number of households with an occupancy of three persons in a three to four person household. CH0! represents the number of households without children*

| Category | Count | Rural | Urban | P3! | P4! | Ch0! | Ch1! | Ch2! | Ch3! |
|---|---|---|---|---|---|---|---|---|---|
| H_P3_A1_€3_EV0_HP0 | 12088 | 12% | 88% | 47% | 53% | 3% | 43% | 53% | 1% |
| H_P3_A2_€3_EV0_HP0 | 35618 | 12% | 88% | 42% | 58% | 3% | 39% | 58% | |
| H_P3_A3_€3_EV0_HP0 | 54445 | 17% | 83% | 37% | 63% | | 35% | 61% | |
| H_P3_A3_€3_EV1_HP0 | 265 | 14% | 86% | 32% | 68% | 3% | 30% | 67% | |
| H_P3_A2_€3_EV0_HP1 | 198 | 35% | 65% | 44% | 56% | | | | |
| H_P3_A3_€3_EV0_HP1 | 635 | 44% | 56% | 37% | 63% | | | | |

The count of each category is a subset of the total number of households in Denmark, as only clean profiles are taken into consideration. Consequently, the count represents the number of profiles present in each category and can be taken as a measure of statistical validity. 'Rural' and 'urban' are indicators of the household´s location, and the percentage represents the share of the count located in those areas. *P3!* and *P4!* show the percentage of households within the group with an occupancy of three or four, whereas *CH0!* to *CH3!* represent the share of children with respectively zero, one or two children below the age of eighteen. Empty fields contain either no numbers or entries that are too low and therefore fail to fulfill GDPR. A category owning both an EV and a HP is not presented, as this combination does not satisfy the GDPR requirements.

**Appendix B**

*B.1. Additional load profile figures and electric vehicle charging occurance*

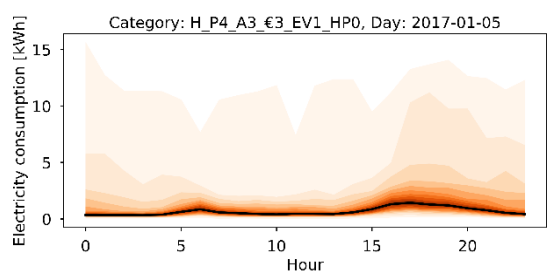
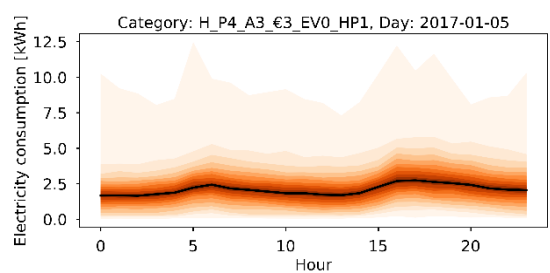

*Figure B11 Electricity consumption of a household living in a house with three to four persons in the large area and income group owning an EV.*

*Figure B12 Electricity consumption of a household living in a house with three to four persons in the large area and income group owning an EV.*

*B.2. Heatmap*

*Table B4 Average hourly consumption of households in kWh without an EV and heat pumps for the entire year.*

| Occupancy | Household | Area | € 1 | € 2 | € 3 |
|---|---|---|---|---|---|
| P1 | Ap | A1 | 0.135 | 0.134 | 0.132 |
| | | A2 | 0.146 | 0.151 | 0.148 |
| | | A3 | 0.173 | 0.188 | 0.214 |
| | H | A1 | 0.209 | 0.237 | 0.249 |
| | | A2 | 0.267 | 0.279 | 0.302 |
| | | A3 | 0.326 | 0.334 | 0.392 |
| P2 | Ap | A1 | 0.165 | 0.175 | 0.165 |
| | | A2 | 0.193 | 0.206 | 0.197 |
| | | A3 | 0.238 | 0.251 | 0.289 |
| | H | A1 | 0.298 | 0.327 | 0.363 |
| | | A2 | 0.372 | 0.371 | 0.397 |
| | | A3 | | 0.424 | 0.466 |
| P3 | Ap | A3 | | | 0.359 |
| | H | A1 | 0.386 | 0.388 | 0.446 |
| | | A2 | 0.472 | 0.445 | 0.476 |
| | | A3 | 0.575 | 0.540 | 0.550 |
| P5+ | H | A1 | 0.499 | 0.481 | 0.532 |
| | | A2 | 0.558 | 0.537 | 0.540 |
| | | A3 | 0.733 | 0.657 | 0.618 |

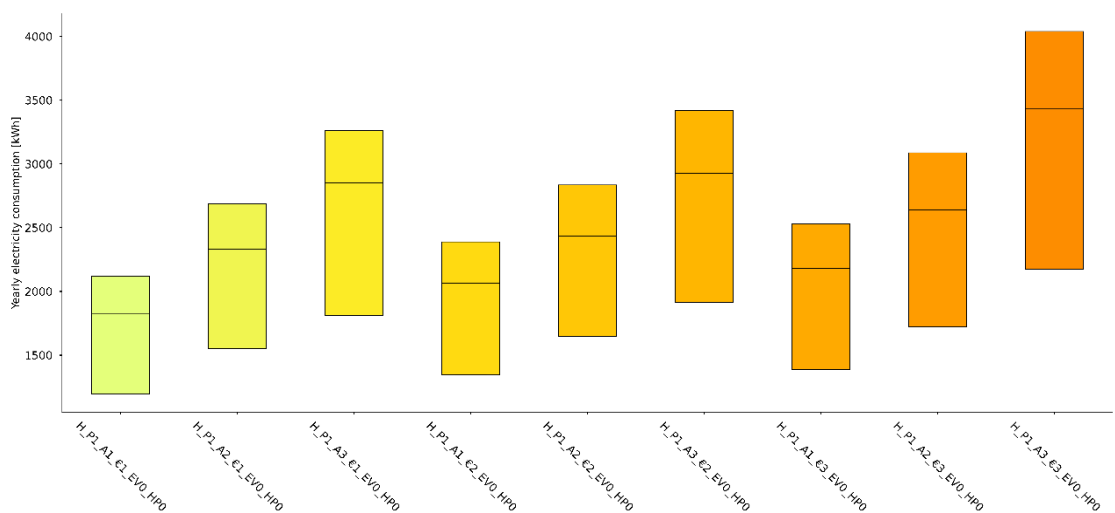

*Figure B13 Annual consumption of households located in apartments. While area influences consumption significantly, income does not show the same pattern, even contrarily. At the same time it has to be noted that the larger the area and income of a one-person apartment the lower the number of households within the group and the more uncertain it is that the actual occupancy might be higher but not registered.*

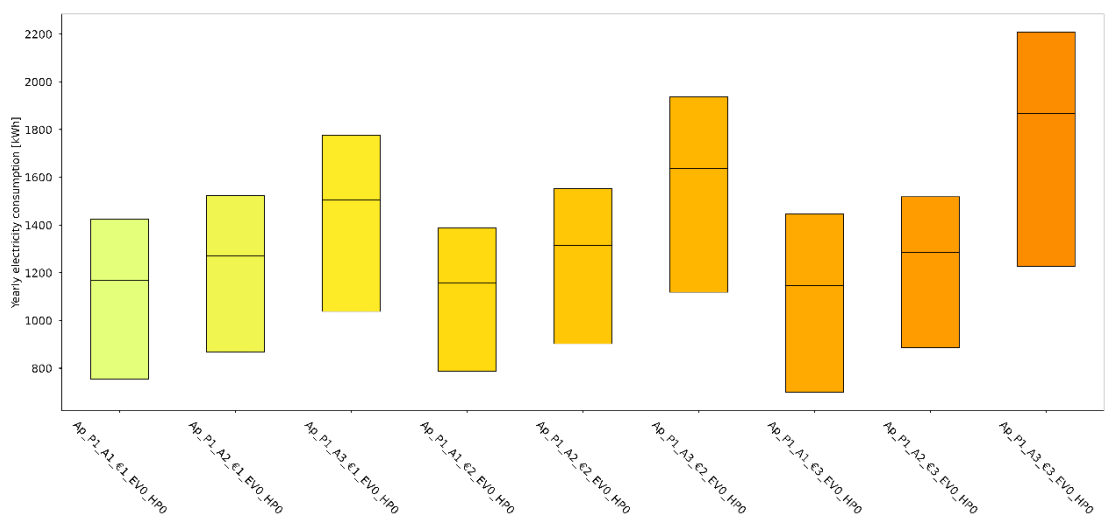

*Figure B14 Annual consumption of households located in a house. Area as well as income show an influence on the annual consumption of the households. The number of one-person households living in a house with a large area and*

*income does not occur that often, and also any discrepancy between registered and actual occupancy might influence the picture.*

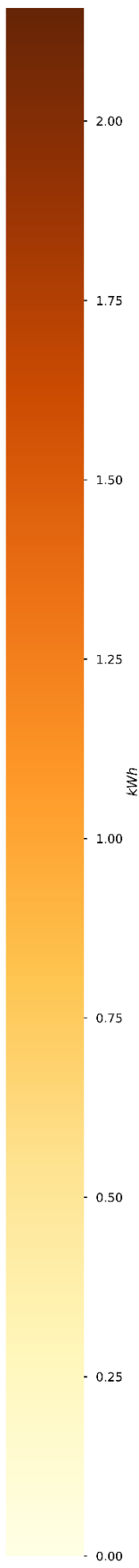

*Figure B15 Average peak hour consumption of consumer category (color) and the variability shown as the standard deviation (number).*

*Table B5 Average mean and variance from 50 random picks of the distributions without and with HP and EV.*

| Peak level | 20% | | 5% | | 1% | |
| --- | --- | --- | --- | --- | --- | --- |
| Technology | noHP | HP | noHP | HP | noHP | HP |
| Av. Mean | 0.69 | 1.62 | 0.76 | 1.94 | 0.98 | 2.23 |
| Av. Var | 0.43 | 1.33 | 0.50 | 1.53 | 0.65 | 1.77 |
| Peak level | 20% | | 5% | | 1% | |
| Technology | noEV | EV | noEV | EV | noEV | EV |
| Av. Mean | 0.69 | 1.10 | 0.76 | 1.19 | 0.98 | 1.56 |
| Av. Var | 0.43 | 2.75 | 0.50 | 2.91 | 0.65 | 4.13 |
| Peak level | 20% | | 5% | | 1% | |
| Technology | HP | EV | HP | EV | HP | EV |
| Av. Mean | 1.62 | 1.10 | 1.94 | 1.19 | 2.23 | 1.56 |
| Av. Var | 1.33 | 2.75 | 1.53 | 2.91 | 1.77 | 4.13 |

*Table B6 Acceptance rate of Welch's t-test with the 0 hypotheses that the mean of one group is equal to the mean of the other group. The first two tests checked the significance of adding a HP and an EV, respectively, compared to households without the technologies. Second, the means of households with HP are checked versus households with EV.*

| Peak level | 20% | | | 5% | | | 1% | | |
| --- | --- | --- | --- | --- | --- | --- | --- | --- | --- |
| Technology | noHP-HP | noEV-EV | HP-EV | noHP-HP | noEV-EV | HP-EV | noHP-HP | noEV-EV | HP-EV |
| Acceptance rate | 0% | 0% | 0% | 0% | 0% | 0% | 0% | 0% | 0% |

*B.3. Load duration curve*

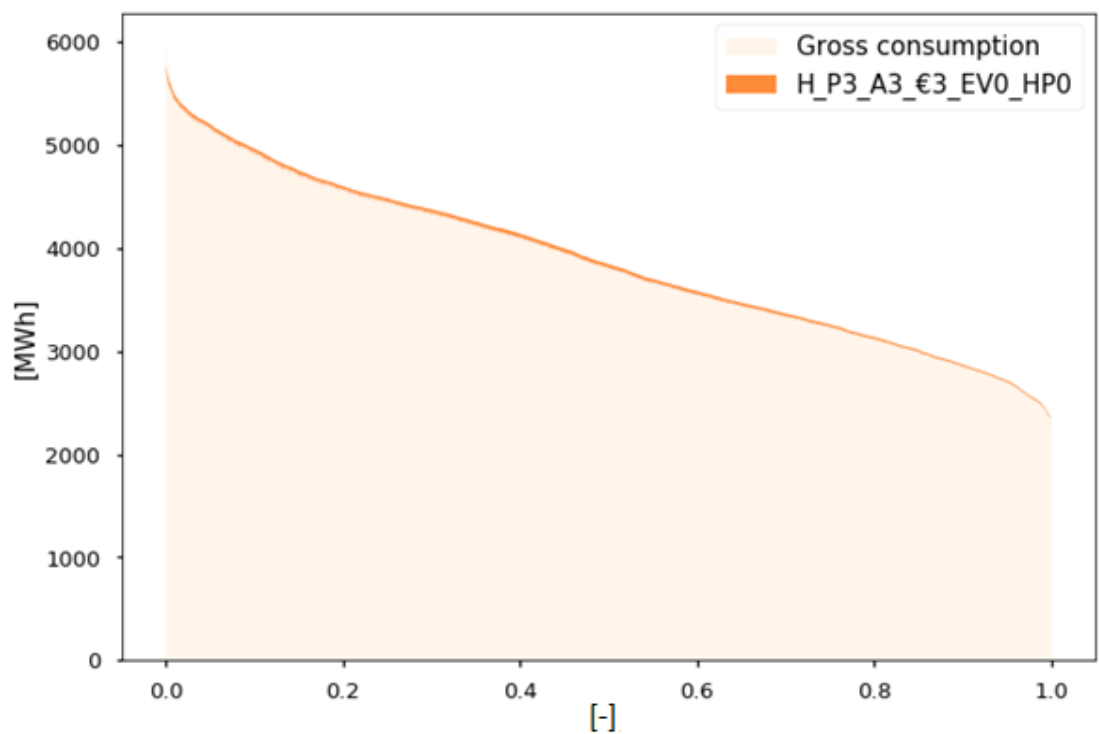

*Figure B16 Load duration curve for 2017 highlighting the share of consumption from the consumer category living in houses with an occupancy of three to four persons, with a large dwelling area and income and without an EV and heat pump. This load duration curve of the consumer category is exchanged and scaled in the study to investigate the impact of EVs and heat pumps.*

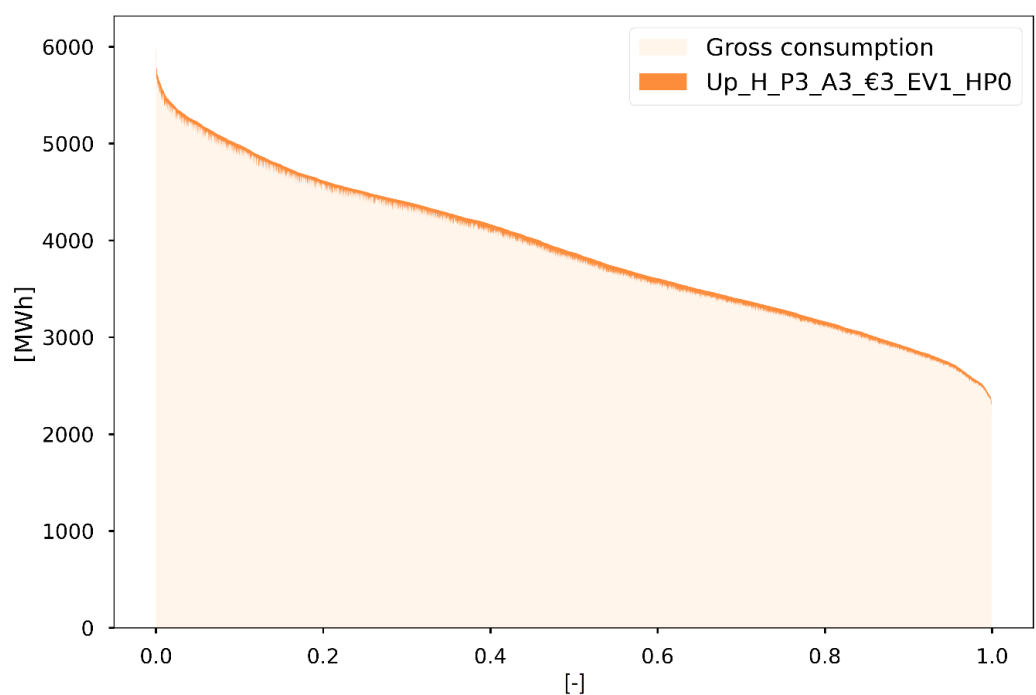

*Figure B17 The Danish load duration curve of 2017 when the entire consumer category of a three to four person household living (P3) in a house without a heat pump (H0) located in the high-income (€3) group and large dwelling area (A3) with 54,445 single households buys an EV. The updated curve converts to "Up_H_P3_A3_€3_EV1_HP0".*

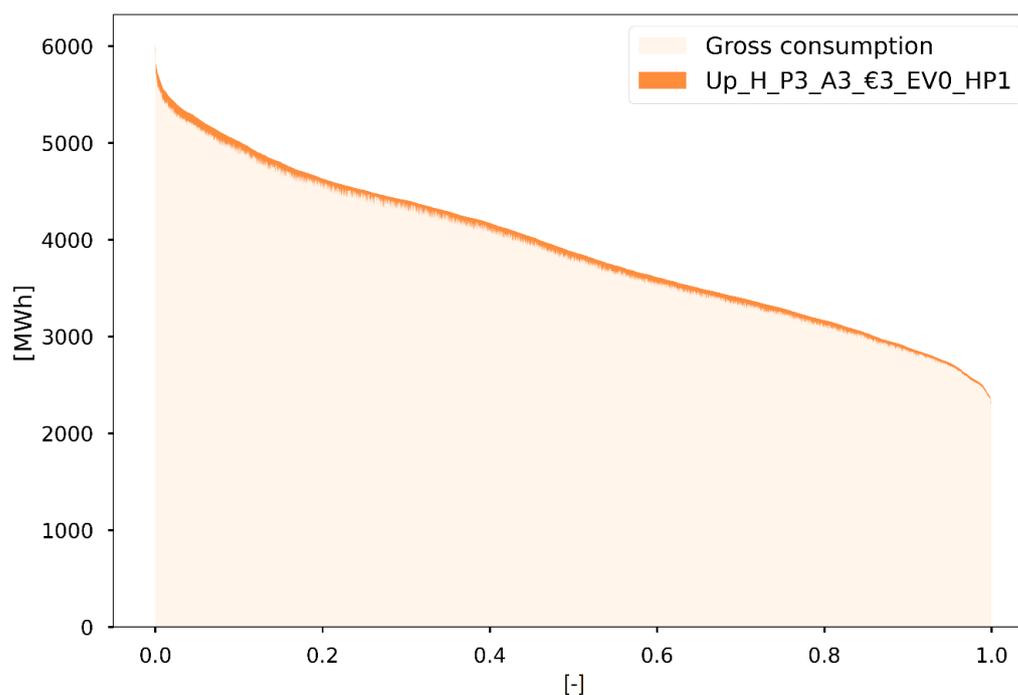

*Figure B18 The Danish load duration curve of 2017 when the entire consumer category of a three to four person household living (P3) in a house without a heat pump (H0) located in the high-income (€3) group and large dwelling area (A3) with 54,445 single households buys a HP. The updated curve converts to "Up_H_P3_A3_€3_EV0_HP1".*

*C.1. The impact of income on residential consumption in Denmark compared to other countries*

Our study further extends the view on the effect of income on electricity consumption. While most studies in e.g., Ireland and the USA show strong effects of income on residential consumption, sometimes being the 2nd most important socio-economic determinant [11,13,51]. Our study shows that the income yields slightly different outcomes from our approach, possibly due to the country-specific characteristics of the Danish case. At the same time, our findings acknowledge that larger income results in a higher probability of owning larger houses, earlier adoption of heat pumps, and ownership of electric vehicles. Our categorization also shows the heterogeneity within the same socio-techno-economic categories, only varying income and not other socio-economic or technology groups. It allows for a more detailed view, revealing that income does not consistently affect electricity consumption anymore. It is not easy to conclude that poorer households have fewer appliances, less efficient appliances, or even a different behavior in Denmark. Contrasting effects in countries such as the USA, where income inequality and the subsequent effects play a more prominent role [6]. The same goes for income effects in developing countries [52].

This adds an interesting aspect for policymaker, grid planning, or city planner. Factors such as dwelling type, size, and the potential amount of occupants are long-term constants compared to the income that can change quickly with occupants moving. At the same time, the local system operator could manage the deployment of EV and HP implies the largest changes in the future of residential electricity consumption.